\documentclass[12pt]{article}
\begin{document}

\begin{titlepage}
\begin{center}
{\Large \bf G\"{o}del Type Metrics in Randall Sundrum Model }

\vspace{5mm}

\end{center}

\vspace{5 mm}

\begin{center}
{\bf Moninder Singh Modgil
\footnote[1]{ Department of Physics, Indian Institute of Technology,
Kanpur, India.}
and Sukanta Panda\footnote[2]{Harish-Chandra Research Institute, India.} }

\vspace{3mm}

\end{center}

\vspace{1cm}

\begin{center}
{\bf Abstract}
\end{center}
Anisotropic cosmological models such as the G\"{o}del universe and
its extensions - G\"{o}del type solutions, are embedded on a
visible 3-brane in the Randall-Sundrum 1 model. The size of the
extra dimension is stabilized by tuning the rotation parameter to
a very small value so that hierarchy problem can be solved. A
limiting case also yields the Randall-Sundrum 2 model. The
rotation parameter on the visible brane turns out to be of order
$10^{-32}$, which implies that visible brane essentially lacks
rotation.

\vspace{1cm}

\end{titlepage}

\section{Introduction}
In last few years, interest in G\"{o}del universe has increased,
following the discovery, that it was an exact solution of the
string theory \cite{Barrow 1998, Kanti 1999, Carrion 1999}\footnote{It is 
also worth mentioning that, in a different
context \cite{goedel1, goedel2}, a whole family of Goedel-like
solutions was investigated by using the AdS/CFT correspondence.}
. Since
its discovery in 1949, G\"{o}del universe attracted attention, due
to unusual properties, such as, existence of Closed Time-like
Curves (CTCs), existence of rotation and consequent anisotropy,
and anti-Machian nature. In 1982 Birch \cite{Birch 1982} argued a
case for observational evidence for universe rotation and
anisotropy. The case for an anisotropic universe, was further
developed by Jain and Ralston \cite{Jain 1999, Ralston 2004}.
G\"{o}del, used a universe density figure of $~10^{-30}$ gm/$cm^3$
and arrived at an angular velocity of $~10^{-18}$ rad/sec. He
considered galaxies as point particle's in a pressureless fluid
permeating the universe. The local vorticity in his model equals
universe's angular velocity. The available data at that time, gave
galaxies an angular velocity $\sim 10^{-14}$. He commented that
his model did not explain random distribution of angular velocity
vectors, and the observed expansion of the universe, which would
modify the predictions. From an observed anisotropy of
polarization of a set of radio source data, Birch \cite{Birch
1982} gave a figure of $10^{-13}$ rad/sec, for angular velocity of
universe. Jain and Ralston have suggested Virgo cluster as the
direction of this anisotropy \cite{Jain 1999, Ralston 2004}.

In an altogether different development in theoretical physics,
Arkani-Hamed, Dimoupulous and Dvali \cite{ADD} proposed a low
scale gravity model in order to solve hierarchy problem by
introducing the concept of large extra dimensions (LED).  Randall
and Sundrum \cite{RS1} suggested introduction of a single extra
dimension in a warped geometry. This had more advantages than ADD
model, from phenomenologically and theoretical view points. It
would be therefore of interest to examine, how an anisotropic
solution, such as the G\"{o}del universe, when used in Randall-
Sundrum (RS) models \cite{RS1, RS2} will behave - primarily as a
check for any signs of anisotropy on the visible brane. In this
paper we come to the conclusion that a visible G\"{o}del type
brane embedded in RS model has vanishingly small rotation.

G\"{o}del universe has the line element,
\begin{equation}
ds^2 = a^2 (dx_0^2 - dx_1^2 - dx_3^2 +
\frac{e^{2x_1}}{2}dx_2^2 +
2e^{x_1} dx_0dx_2) \label{Godel line element}
\end{equation}
which satisfies the Einstein equations for a uniform matter
density, $\rho = (8 \pi G a^2)^{-1}$, zero pressure, and a
cosmological constant, $\Lambda = -4 \pi G \rho$ (here $G$ in
Newton's gravitational constant). The corresponding angular
velocity is: $\omega = 2(\pi G \rho)^{1/2} = (\sqrt2 a)^{-1}$. Any
circle in $(x_1 x_2)$-plane with radius $R_c=c/\omega$ (where $c$
is speed of light) is a closed null curve (CNC). It acts as a
horizon for particles starting from its center and moving along
geodesics. Its interesting to compare G\"{o}del horizon with
Friedmann-Robertson-Walker (FRW) horizon. Whereas FRW is expanding
at speed of light, G\"{o}del horizon is rotating at speed of
light. Circles of radii larger than that of CNC, are Closed
Time-like Curves (CTCs). For very small universe densities, radii
of the CTCs will be very large, and presence of CTCs will largely
be immaterial to local physics.

If in Einstein's field equations for G\"{o}del universe, one puts
cosmological constant equal to zero, then one obtains a solution
with non-zero pressure. The two situations may be expressed as,
\begin{equation}
\begin{array}{cc}
 8\pi G p_{matter}=8\pi G \rho_{matter}=\frac{a^2}{2},
& \lambda_{cosmological}=0, or \\
  8 \pi G\rho_{matter}=a^2, p_{matter}=0, &  8 \pi G
\lambda_{cosmological}=-\frac{a^2}{2}.
\end{array}
\end{equation}

The line element of Randall Sundram (RS) models,
\begin{equation}
ds^2 = e^{-2\sigma(y)}\eta_{\mu\nu}dx^{\mu}dx^{\nu}+ d
y^2 \,
\end{equation}
has a structural compatibility with G\"{o}del line element, in the
sense that the pre-factor $e^{-2\sigma(y)}$, can be directly
equated with pre-factor $a^2$ in G\"{o}del line element. This
allows exponential variation of rotation, pressure and density in
bulk direction, which makes the fluid anisotropic in bulk
direction. This however, is not unusual in physics. For example,
earth's atmosphere has gradients of pressure, density etc., in
vertical direction, which counter balance earth's gravitational
pull. In RS models particles gravitate from hidden brane towards
visible brane. In the G\"{o}del type branes embedded in RS models,
pressure gradient in the bulk direction, of the 5-dimensional
anisotropic fluid, provides an additional force, which pushes
particles towards visible brane.

The solution to Einstein equations in presence of a 5-D bulk
cosmological constant $\Lambda$ in RS set-up is given by
\begin{equation}
\sigma(y)=\sqrt{-\frac{\Lambda}{24 M^{3}}}~ |y| = k |y|
\end{equation}
This solution is valid only if brane tensions and cosmological constants
are related as follows
\begin{equation}
\Lambda = -24 M^{3}k^2,~~v_1=-v_2=24 M^{3}k
\label{cons}
\end{equation}
If $k > 0$, the brane 1 at $y=0$ has positive tension $v_1$ and
brane 2 at $y =  y_{c} $ has negative tension $v_2$. Starting from
the 5-D action, after integrating over extra coordinate $y$ yields
the reduced effective 4-D plank scale given by this relation
\begin{equation}
M_{Pl}^2 = \frac{M^3}{k}(1-e^{-2k y_c})
\end{equation}
Setting the visible brane at $y=y_c$ where our standard
model fields live, it is found that any mass parameter $m_0$ on the
visible 3-brane in the fundamental higher-dimensional theory will
correspond to a physical mass $$m=e^{- k y_c}m_0$$
TeV mass scales can be generated on
the 3-brane at $y=y_c$ due to the exponential factor present in the
metric which fixes $ky_c \approx 35$. Next section we construct solutions to
Einstein equations in RS set-up in presence of bulk fluid.

\section{Solution In Presence of Anisotropic Bulk Fluid}
Consider a metric given by,
\begin{equation}
ds^2 = e^{-2 \sigma(y)} {\hat g}_{\alpha \beta}(x) dx^{\alpha}
dx^{\beta} + dy^2 \ ,
\label{metric}
\end{equation}
where $\hat{g}_{\alpha \beta }$ is the induced 4-D metric. By
assuming that the bulk matter has no flow along y-direction, i.e.,
$u^5=0,$ where, $u^5$ is the velocity vector along y-direction, we
arrive at the following form for 5-D anisotropic energy momentum
tensor of bulk fluid,
\begin{equation}
T^{AB} =\left(
\begin{array}{ll}
T^{\alpha \beta } & 0 \\
0 & P%
\end{array}%
\right) \;,
\end{equation}
where $T^{\alpha \beta }$ is energy-momentum tensor of the 4-D
perfect fluid, given by,
\begin{equation}
T^{ab} = A u^{a} u^{b} + B g^{ab}
\end{equation}
and
\begin{equation}
T^{33} = C g^{33} \ ,
\end{equation}
where $a,b$ run over $0,1,2$. From the energy momentum
conservation $T^{5B};_{B}=0$ we obtain the condition
\begin{equation}
P^{\prime }=-2 \sigma^{\prime}(y) (-A + 3B + C - 4P) \ ,
\end{equation}
where prime stands for partial derivative with respect to $y$. The
above equation is referred to as the 'hydrostatic equilibrium
equation' of the bulk fluid, along the $y$-direction \cite{lp}.

In RS model, $5$-D Einstein equations are given by
\begin{equation}
R_{AB} - \frac{1}{2} g_{AB} R= \frac{1}{4 M^3}(T_{AB}
-\lambda g_{AB} - v_1 \delta(y)- v_2 \delta(y-y_c)) \ .
\label{five}
\end{equation}
Substituting the metric Eq. (\ref{metric}) into Eq. (\ref{five})
and using the relation Eq. (\ref{cons}), we get the $4$-D Einstein
equations,
\begin{equation}
R^{(4)}_{\alpha \beta} - \frac{1}{2} g_{\alpha \beta} R^{(4)} =
\frac{1}{4 M^3} T_{\alpha \beta}
\label{r4}
\end{equation}
where the four dimensional Ricci scalar is given as
\begin{equation}
R^{(4)} = -\frac{1}{2 M^3} P e^{-2 \sigma(y)} \ .
\label{r41}
\end{equation}
As can be seen right hand side of equations (\ref{r4}) and
(\ref{r41}) depend upon the coordinate $y$ which implies that
these are not four dimensional tensors. However by relating 4-D
energy momentum tensor to 5-D energy momentum tensor, the
effective 4-D Einstein equation on the visible brane in presence
of anisotropic perfect fluid source, can be written as,
\begin{equation}
R^{(4)}_{ab} - \frac{1}{2} g_{ab} R^{(4)} = 8 \pi G_4 T^{(4)}_{ab} \ ,
\end{equation}
\begin{equation}
R^{(4)}_{33} - \frac{1}{2} g_{33} R^{(4)} = 8 \pi G_4 T^{(4)}_{33} \ ,
\end{equation}
where $T^{(4)}_{ab}$ and $T^{(4)}_{33}$ are given by,
\begin{equation}
T^{(4)}_{ab} = {\hat A} {\hat u}_{a} {\hat u}_{b} + {\hat B} {\hat g}_{ab}
\end{equation}
and
\begin{equation}
T^{(4)}_{33} = {\hat C} {\hat g}_{33} \ ,
\end{equation}
where
\begin{equation}
A = c_0 e^{2 \sigma(y)} {\hat A}, ~~~B = c_0 e^{2 \sigma(y)}{\hat
B}, ~~~C = c_0 e^{2 \sigma(y)} {\hat C} \,
\end{equation}
\begin{equation}
u_{a} =e^{\sigma(y)} {\hat u}_{a}
\end{equation}
and $c_0= 32 M^3 \pi G_4.$ $A, B$ and $C$ satisfy the relation
\begin{equation}
2 P =-A + 3 B + C \ .
\end{equation}

\section{G\"{o}del Type Solutions On The Visible brane}
As indicated in the previous section, one can embed any solution
of the four dimensional Einstein equations with perfect fluid
source, in RS models. One of option is the G\"{o}del solution,
\cite{godel} which describes an anisotropic, rotating universe. We
also consider other G\"{o}del type solutions, which can give rise
to a more realistic case. Basic idea is to show that if we are
living in a 3-brane, then it should represent a universe with
negligible rotation, even though there may be significant rotation
in bulk and on the hidden brane.

{\bf Case A:} For $A = \rho,$ $B = 0$ and $ C=0$ the solution is,
\begin{equation}
ds^2= e^{-2 k |y|}\left(a^2(dt^2-dx_1^2+(e^{2x_1}/2)
dx_2^2-dx_3^2+2e^{x_1}dt dx_2)\right) + dy^2,
\end{equation}
However this solution is not stable in this case, as proved in
next section.

{\bf Case B:} Consider a expanding, rotating and shear free
universe as given in Ref. \cite{gron} on the brane with equation
of state $\hat p=-\hat{\rho}.$ This solution gives rise to
inflationary scenario in presence of rotation. In ref.
\cite{gron}, it is shown that vorticity decreases exponentially as
universe expands exponentially. If we embed this universe on the
brane, the solution takes the form,
\begin{equation}
ds^2 = e^{-2 k |y|} [(dt + A(t) w^1)^2 -
B^2(t) ((w^1)^2+ (w^2)^2 + (w^3)^2)] + dy^2 \,
\end{equation}
where expansion is generated by time dependent scale factors
$A(t)$ and $B(t)$,  and $w^1, w^2, w^3$ are one forms
\begin{eqnarray}
w^1 &=& - \sin x^1 dx^1 + \sin x^1 \cos x^3 dx^2 , \\
w^2 &=&   \cos x^3 dx^1 + \sin x^1 \sin x^3 dx^2, \\
w^3 &=&   \cos x^1 dx^1 + dx^3.
\end{eqnarray}
Effective $4$-D Einstein equation is given by,
\begin{equation}
R^{(4)}_{\mu \nu}=- \Lambda {\hat{g}}_{\mu \nu},
\end{equation}
where $\Lambda$ is the $4$-D cosmological constant. Solution of
Einstein equation, gives \cite{gron}
\begin{equation}
A(t)= A_0 \sinh^{1/2}(H_0 t) sech(H_0 t)
\end{equation}
and
\begin{equation}
B(t)= \frac{1}{2 H_0} \cosh(H_0 t),
\end{equation}
where $H_0 = \sqrt{\frac{\Lambda}{3}}$ and $A_0$ is a constant.
Vorticity is defined by \cite{gron}
\begin{equation}
V^\mu = -\frac{1}{2} \eta^{\mu \nu \alpha \beta} u_{\nu;\alpha}
u_{\beta},
\end{equation}
where $u^\mu$ is the four velocity vector of the cosmic fluid,
$u_{\nu;\alpha}$ is the covariant derivative of the four velocity
vector of the cosmic fluid and $\eta$ is the completely
antisymmetric tensor. Expansion rate is given by \cite{gron}
\begin{equation}
H=3 \frac{\dot{B}}{B} = 3 H_0 \tanh(H_0 t), \label{expa}
\end{equation}
and first component of vorticity is
\begin{equation}
V^1 = \frac{1}{2} \frac{A}{B^2}.
\label{vor}
\end{equation}

In next section we show that stabilization of extra dimension sets
the time scale at which inflation takes place. Duration of
inflation determines the decay in vorticity.

{\bf Case C:} Static G\"{o}del matric in the presence of perfect
fluid source other than $p=0$ G\"{o}del solution derived in
\cite{romano} can be embedded on the 3-brane. Solution is given by
in terms cylindrical coordinates
\begin{equation}
ds^2= e^{-2 k |y|}\left(dt^2+\mu^2 (\gamma-1)^2d\phi^2 - \sigma^2 d\phi^2
-dr^2-dx_3^2+ 2 \mu(\gamma-1)dt d\phi\right) + dy^2,
\end{equation}
where $\gamma= \cosh r$, $\sigma=\sinh r$, and $\mu$ is the
rotation parameter in the theory. These solutions represent a
rotating universe in presence of electromagnetically charged
perfect fluids for $\mu^2 > 2$ \cite{romano}. In the next section
we will derive the constraint on $\mu$ by stabilizing the size of
the extra dimension.

\section{Stabilization of the Extra Dimension}
Stabilization of size of the extra dimension can be achieved by
minimizing the effective four dimensional curvature on the brane,
by integrating over extra dimension. The five dimensional
curvature is given by
\begin{equation}
R = - \frac{1}{6 M^3} \left(g^{AB}T_{AB} -5 \Lambda
- 4 (v_1 \delta(y)+v_2 \delta(y-y_c))\right)
\label{curv}
\end{equation}
For the general metric (\ref{metric})
\begin{equation}
g^{AB}T_{AB}= -A+3B+C+P= 3P
\label{trace}
\end{equation}
Five dimensional action is given by
\begin{equation}
S_{5D} = \int d^4x \int_{-y_c}^{y_c} 2 M^3 \sqrt{-g} R dy
\label{action}
\end{equation}
Substituting Eq. (\ref{curv}) and Eq. (\ref{trace}) into Eq.
(\ref{action}) and after integrating over $y$-coordinate, we get
\begin{equation}
S_{5D} = \int d^4x\sqrt{-\tilde{g}} \frac{-1}{3} \left(-\frac{6 M^3}{k}
R_{(4)} \left(1- e^{-2 k y_c}\right) -\frac{5}{2k} \Lambda (1- e^{-4ky_c})
-4 v_1-4 v_2 e^{- 4 ky_c}\right)
\label{action1}
\end{equation}
Using the relation among $\Lambda, v_1$ and $v_2$, effective
four dimensional curvature becomes
\begin{equation}
{\tilde{R}}_{(4)} = \frac{6 M^3}{k}
R_{(4)} \left(1- e^{-2 k y_c}\right) + 36 M^3 k (1- e^{-4ky_c}).
\label{eff}
\end{equation}
${\tilde{R}}_{(4)}$ is the effective four dimensional curvature.
One can easily see from the Eq. (\ref{eff}) that effective 4-D
curvature has a minimum only when $R_{(4)}$ becomes negative. By
differentiating with respect to $y_c$ we get
\begin{equation}
e^{-2 k y_c} = - \frac{R_{(4)}}{12 k^2}.
\label{size}
\end{equation}

{\bf Case A:} In this case $R_{(4)} = 1/a^2$ is positive and
${\tilde{R}}_{(4)}$ does not have a minimum.

{\bf Case B:} In this case $R_{(4)} = 4 \lambda.$ Eq. (\ref{size})
leads to
\begin{equation}
\Lambda=3 k^2 e^{-2 k y_c}.
\end{equation}
If we take $\Lambda=3 k^2 \times 10^{-32}$ then we get $ky_c
\approx 40$ which solves hierarchy between plank scale and
electroweak scale. If we take $k=M_{pl}, \Lambda \approx 10^{-60}
sec^2$, than inflation should occur at $10^{-30}$ second in
contrast to GUT time $10^{-35}$ second. Expansion rate given in
Eq. (\ref{expa}) corresponds to de-Sitter universe. Inflation ends
at $\sim 10^{-28}$ seconds in our case. During inflationary period
vorticity decreases by an order of $10^{-55}$ which implies that
effects of rotation are diluted by inflation. This is effectively
a transition from a rotating and expanding universe, to an
expanding universe with little rotation.

{\bf Case C:}
  In this case four dimensional curvature is
\begin{equation}
R_{(4)} = - \frac{(\mu^2 - 4)}{16 \pi G_4}.
\end{equation}
If $\mu^2 = 4,$ we obtain a RS2 model \cite{RS2} where $y_c \rightarrow
\infty,$ i.e. infinite fifth dimension.
If $\mu^2 > 4, $ relation between the rotation parameter
and the size of the extra dimension becomes,
\begin{equation}
e^{- 2 k y_c} = \frac{(\mu^2 - 4)}{8 \pi G_4}\frac{1}{24 k^2}.
\end{equation}
If $\frac{(\mu^2 - 4)}{8 \pi G_4}\frac{1}{24 k^2} = 10^{-32},$
then we get $ky_c\approx40$, which is the value needed to solve
the hierarchy problem. This value is also obtained by another
stabilization mechanism \cite{gw}.

\section{Discussion}
For $k\sim M_{pl},$ the term $(\mu^2 - 4)$, which determines the
strength of the rotation in this universe (Case C), turns out to
be order of $10^{-32}$. This is too small to show any appreciable
effect of rotation on the visible brane. The vanishingly small
value of rotation parameter as calculated for RS model in this
paper, and other estimates of angular velocity (indicated in
introduction) therefore suggest the conclusion, that if our
universe lives on a 3-brane explained by RS1 model then universe
must be a non-rotating one, without any anisotropy. On the other
hand, if the anisotropic models are correct, than RS models must
require additional input. Detailed investigation also needs to be
done in case of a rotating and expanding G\"{o}del type universe
\cite{Obukhov 2000} embedded as a brane in RS models.

\end{document}